\documentclass[reprint,prl,showpacs,superscriptaddress]{revtex4-2}
\usepackage{amssymb}
\usepackage{amsmath}    
\usepackage{graphicx}   
\usepackage{verbatim}

\usepackage{lineno}
\usepackage{color}      
\usepackage{booktabs} 

\begin{document}

\title{
Relativistic plasma aperture for laser intensity enhancement
}

\author{M. Jirka}
\affiliation{ELI Beamlines Centre, Institute of Physics, Czech Academy of Sciences, Za Radnici 835, 25241 Dolni Brezany, Czech Republic}
\affiliation{Faculty of Nuclear Sciences and Physical Engineering, Czech
Technical University in Prague, Brehova 7, 115 19 Prague, Czech Republic}

\author{O. Klimo}
\affiliation{ELI Beamlines Centre, Institute of Physics, Czech Academy of Sciences, Za Radnici 835, 25241 Dolni Brezany, Czech Republic}
\affiliation{Faculty of Nuclear Sciences and Physical Engineering, Czech
Technical University in Prague, Brehova 7, 115 19 Prague, Czech Republic}

\author{M. Matys}
\affiliation{ELI Beamlines Centre, Institute of Physics, Czech Academy of Sciences, Za Radnici 835, 25241 Dolni Brezany, Czech Republic}
\affiliation{Faculty of Nuclear Sciences and Physical Engineering, Czech
Technical University in Prague, Brehova 7, 115 19 Prague, Czech Republic}

\begin{abstract}
A substantial increase in local laser intensity is observed in the near field behind a plasma shutter. 
This increase is caused by the interference of the diffracted light at the relativistic plasma aperture and it is studied both analytically and using numerical simulations.
This effect is only accessible in the regime of relativistically induced transparency and thus it requires a careful choice of laser and target parameters.
The theoretical estimates for the maximum field strength and its spatial location as a function of target and laser parameters are provided and compared with simulation results.
Our full 3D Particle-in-Cell simulations indicate that the laser intensity may be increased roughly by an order of magnitude improving the feasibility of strong field QED research with the present generation of lasers.
\end{abstract}

\maketitle
\section{Introduction}
Nowadays, the most powerful laser systems can reach intensity on the order $ 10^{22}~\mathrm{W/cm^{2}} $ and the dynamics of laser-plasma interaction becomes strongly relativistic \cite{Bahk2004,Yanovsky2008,Pirozhkov2017}.
One example is the onset of relativistic transparency in the interaction with the over-dense target \cite{Fernndez2017}.
The laser light cannot propagate in the over-dense target, as its frequency is lower or equal to the frequency of the electron plasma oscillations.
However, when the intense laser field is applied, the electron mass increases due to the relativistic motion reducing the plasma frequency.
It results in a decrease of the effective target density and thus the intense part of the laser field can propagate through the plasma.

Such an effect is especially important in the interaction with a thin over-dense foil, as it affects the temporal envelope of the laser pulse \cite{Vshivkov1998,Reed2009,Palaniyappan2012,Wei2017}.
%
Therefore, it has an impact on ion acceleration and radiation generation \cite{Henig2009,Kiefer2013}.
%
It was shown that in the case of an ultra-intense laser pulse, a relativistic plasma aperture is created in an ultra-thin foil and the laser pulse is diffracted \cite{GonzalezIzquierdo2016Optically}.
Since the plasma electrons are driven by the diffraction pattern of the laser field, the spatial structure of the accelerated electron beam can be controlled by varying the laser pulse parameters \cite{GonzalezIzquierdo2016Optically}.
As the electron dynamics consequently affects the spatial-intensity distribution of the accelerated ions, such an approach could be used for optically controlled ion acceleration \cite{GonzalezIzquierdo2016Towards,Higginson2018,Williamson2020}.

Here we study the increase of the laser field intensity caused by the interference of the light diffracted at the relativistic plasma aperture for the case of linearly polarized laser pulse.
It is investigated both analytically and using 2D and 3D Particle-In-Cell (PIC) simulations.
It is shown that due to the relativistic motion of the electrons, the laser intensity can be locally increased by almost an order of magnitude.
However, this effect is only accessible in the regime of relativistically induced transparency and thus it requires a careful choice of the laser and target parameters.
If the target is either under- or over-dense, the relativistic plasma aperture cannot be created and the diffraction structure responsible for the field intensity enhancement is not present.
The theoretical estimates for the maximum field strength and its spatial location are provided as a function of laser pulse and target parameters and compared with simulation results.
%

\section{Interaction with a plasma shutter}
%
In the following text, we study the process of laser intensity enhancement in the interaction of an ultra-intense laser pulse $( \sim10^{22}~\mathrm{W/cm^{2}} )$ with an ultra-thin $(\sim 10\mathrm{s}~\mathrm{nm})$ plasma layer having the electron plasma density $ n_{\mathrm{p}} $ greater than the value of the relativistic critical density $ n_{\mathrm{c}\gamma}=\gamma n_{\mathrm{c}} $ where $ \gamma $ is the relativistic Lorentz factor (of an electron having the charge $ e $ and mass $ m_{e} $) and $ n_\mathrm{c}= \omega_{0}^{2}m_{e}/\left(4\pi e^{2} \right) $ represents the non-relativistic critical plasma density.
%
%
We assume a laser pulse linearly polarized along the $ y $-axis propagating in the negative $ x $-direction focused to a spot of radius $ w_{0} $.
It has a Gaussian temporal envelope of full width at half maximum (FWHM) duration $ \tau $ in the laser intensity.
Its amplitude is characterized by the Lorentz invariant parameter $ a_{0}=eE_{0}/\left(m_{e}\omega_{0}c \right)  $ where $ E_{0} $ is the amplitude
of the electric field, $ \omega_{0} $ is the laser angular frequency and $ c $ is the speed of light \cite{Gibbon2005}.
%

When a relativistically over-dense shutter ($ n_{\mathrm{p}}>n_{\mathrm{c\gamma}} $) is assumed, the front part of the laser pulse is reflected from the foil surface.
However, as the laser pulse intensity grows to its maximum value, the laser pulse starts to penetrate the target.
The dominant mechanism allowing target penetration differs depending on the target thickness and density.
In the case of the ultra-thin foil, the electrons are pushed away due to the radiation pressure of the incoming laser pulse.
If the thicker target is assumed, it is rather the increasing relativistic mass of electrons that causes the foil to become relativistically transparent so that the rest of the laser pulse can pass through the foil.
Therefore, the laser field strength at which the plasma layer becomes transparent depends on the density and thickness of the foil as well as on the laser wavelength $ \lambda $.
%
%
In both cases, the laser pulse acquires a steep front edge as a result of the interaction \cite{Vshivkov1998,Reed2009,Palaniyappan2012}, that can be utilized, e.g., for enhanced photon emission due to the improved temporal contrast or for enhancement of ion acceleration from structured targets due to reduced development of transverse short-wavelength instabilities \cite{Jirka2020,Matys2020}. 
Here we study the interaction with the ultra-thin foil.
For $ a_{0}\gg1 $  and an unperturbed plasma slab of thickness $ l \ll \lambda $, the laser pulse strength required for penetration of the foil (meaning that the laser pulse is able to push all electrons away from the initial position of the foil by the radiation pressure) can be estimated as \cite{Vshivkov1998,Bulanov2012,Bulanov2016}
\begin{equation}\label{a0reqMax}
a_{0}^{\mathrm{max}}=\dfrac{n_{\mathrm{p}}\pi l}{n_{\mathrm{c}}\lambda}.
\end{equation}

Naturally, the laser pulse penetrates the target near the laser axis at first.
With the increase of the laser pulse strength at later times, the target becomes transparent at all points where the condition \eqref{a0reqMax} for laser penetration is satisfied.

However, even for $  a_{0} < a_{0}^{\mathrm{max}}$, the laser pulse may penetrate through the foil at least partially due to the skin effect \cite{Gamaliy1990}.
The relativistically corrected skin depth is $ l_{\mathrm{skin}}=\sqrt{\gamma}c/\omega_{\mathrm{p}} $, where $ \omega_{\mathrm{p}}=\sqrt{4\pi n_{\mathrm{p}}e^{2}/m_{e} } $ is the plasma frequency \cite{Gibbon2005}.
The radiation pressure of the laser pulse will first squeeze the plasma layer of initial thickness $ l $ to a thin slab 
of thickness $ d $ while the density will grow by the factor of $ l/d $.
The corresponding skin depth is $ c/\omega_{0}\sqrt{\gamma 
n_{\mathrm{c}}d/(n_{\mathrm{p}}l)} $.
As is shown later in the text, the amplitude of the resulting field is enhanced due to the interference of the transmitted and scattered beams.
Since the strength of the scattered beam should be non-negligible in order to noticeably contribute to the interference, we consider the target to be transparent if the intensity of the transmitted (scattered) beam is at least of the order of $ \exp\left( -3/2\right)  $ of the incident intensity.
Thus, assuming the intensity of the incident laser pulse is reduced by a factor of $ \exp\left( -3/2\right)  $ after passing through the squeezed plasma layer of a thickness $ d $, we obtain
\begin{equation}\label{intensity_drop}
2k_{0}\sqrt{\dfrac{ldn_{\mathrm{p}}}{\gamma 
n_{\mathrm{c}}}}=\dfrac{3}{2},
\end{equation}
where $ k_{0}=\omega_{0}/c $.
After finding the value of $ d $ from Eq.~\eqref{intensity_drop} and inserting into $ a_{0}^{\mathrm{t}}=n_{\mathrm{p}}\pi \left( l-d\right) /n_{\mathrm{c}}\lambda $ one may obtain the characteristic amplitude of the field required for penetration of the target
%
\begin{equation}\label{a0reqt}
a_{0}^{\mathrm{t}}=\dfrac{4/3\sqrt{2}k_{0}^{2}l^{2}n_{\mathrm{p}}}{\left( 3/4+8/3\sqrt{2}k_{0}l\right) n_{\mathrm{c}}}.
\end{equation}
%

%

\begin{figure}
\centering
\includegraphics[width=1.0\linewidth]{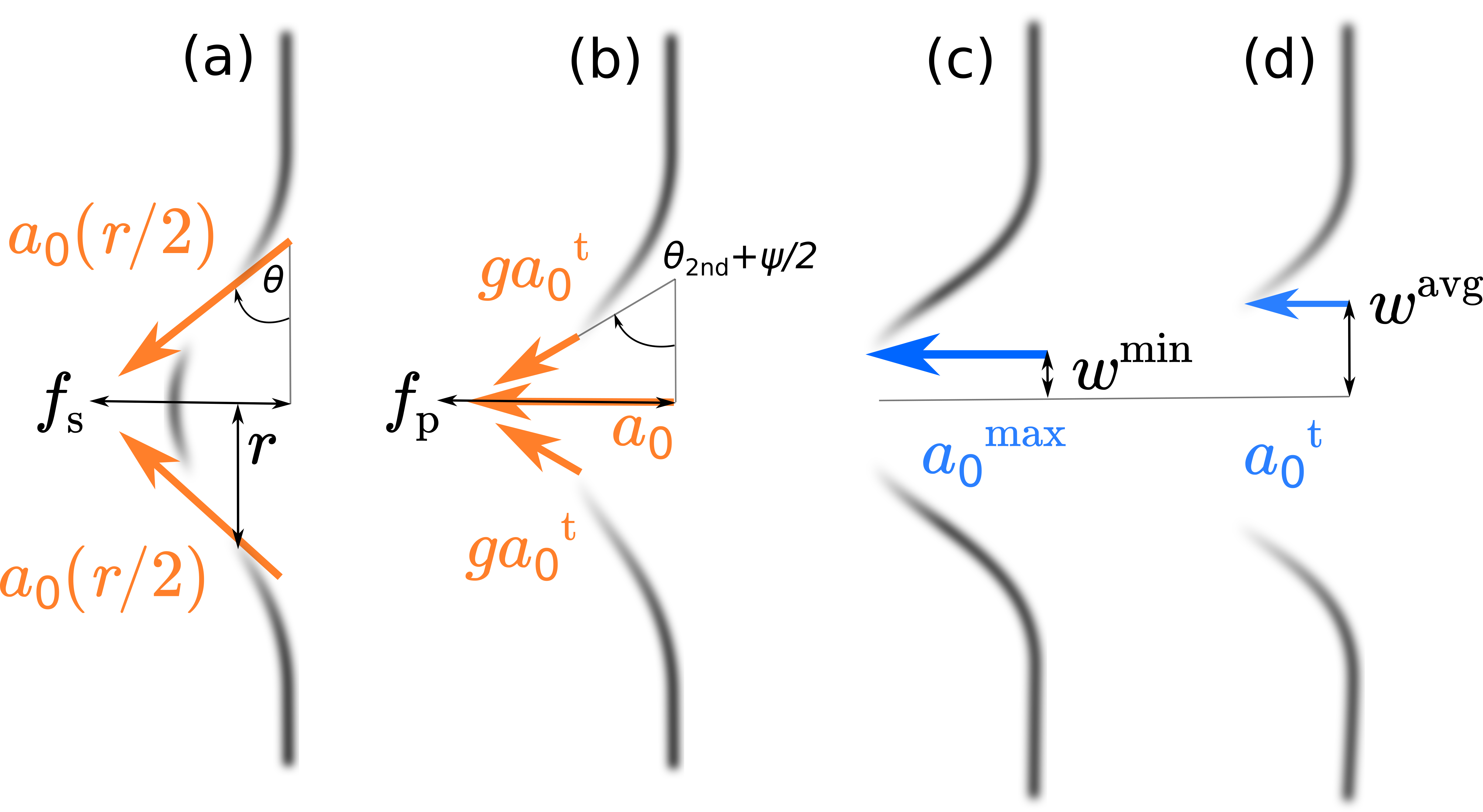}
\caption{Schematic of target geometry during the interaction with (a) s-polarized and (b) p-polarized laser pulse. (c) Minimal and (d) characteristic aperture radius for penetration of the foil.}
\label{fig:schematic}
\end{figure}

For the ultra-thin foils with the thickness of the order of 10 nm assumed in this paper, $ a_{0}^{\mathrm{t}} \approx 0.5a_{0}^{\mathrm{max}} $.
While the most intense part of the laser pulse propagates through the thin plasma slab and a relativistic aperture is formed the electrons at the borders of the aperture oscillate in the direction of the resulting electric field, which is given by the interference of the incident laser wave with the wavefronts reflected from relativistically overdense plasma slab \cite{GonzalezIzquierdo2016Optically}.
In the case of s-polarization, target electrons are simultaneously driven towards the laser beam axis by the ponderomotive force of the two nodes created at the target front side twice per laser period \cite{GonzalezIzquierdo2016Optically}.
It results in an increase of the target density on the axis and therefore two separate relativistic plasma apertures are created in the vicinity of the laser beam axis at first, see Fig.~\ref{fig:schematic}(a) \cite{GonzalezIzquierdo2016Optically}.
It agrees with a double lobe profile of the transmitted laser field observed in three-dimensional simulations and the corresponding asymmetry measured in the distribution of accelerated electrons and ions \cite{Yin2011,Gray2014,GonzalezIzquierdo2016Towards,GonzalezIzquierdo2016Optically}.
We assume that at the radial position $ r $, i.e., at the outer edge of the relativistic plasma aperture, the laser pulse strength is equal to $ a_{0}^{\mathrm{t}}=a_{0}\exp\left( -r^{2}/w_{0}^{2} \right)  $.
As the high density along the beam axis prevents the transmission of the beam close to the axis, the strength of the field passing through one aperture can be estimated as $ a(r/2)=a_{0}\exp\left[ -r^{2}/\left( 2w_{0}\right)^{2} \right]  $.
The resulting field at the rear side of the target is created by the interference of two transmitted beams, each having the above mentioned strength.
Therefore, the amplitude of the diffracted field is approximately given by 
\begin{equation}\label{a0resS}
a_{0}^{\mathrm{s}}=2a(r/2)=2a_{0}\exp\left[ -r^{2}/\left( 2w_{0}\right)^{2} \right].
\end{equation}

When p-polarization is assumed, the aperture is created on the axis at first, see Fig.~\ref{fig:schematic}(b).
Contrary to the previous case, dense electron bunches are driven alternatively from two opposite edges of the aperture towards the laser axis at the frequency $ 2\omega_{0} $  \cite{Duff2020}.
The photons of the transmitted laser pulse are thus scattered by these relativistic electron bunches generating a number of high harmonics.
The intensity of the $ n $th harmonic can be approximated by a power-law $ I_{n}\propto n^{-8/3} $ \cite{Baeva2006,Edwards2020}.
Assuming that laser field of strength $ a_{0}^{\mathrm{t}} $ is scattered by the electron bunches at two opposite sides of the aperture, the interference of the generated harmonics will create a field of strength $ 2g a_{0}^{\mathrm{t}} $, where $ g=\sqrt{\sum_{n=2}^{N}n^{-8/3} } $ is the fraction of the incoming field scattered into $ N $ harmonics.
Due to their further interference with the field propagating along the laser axis, the resulting field amplitude can be approximated as
\begin{equation}\label{a0resP}
a_{0}^{\mathrm{p}}= 2g a_{0}^{\mathrm{t}}+a_{0}.
\end{equation}
\begin{figure}
\centering
\includegraphics[width=1.0\linewidth]{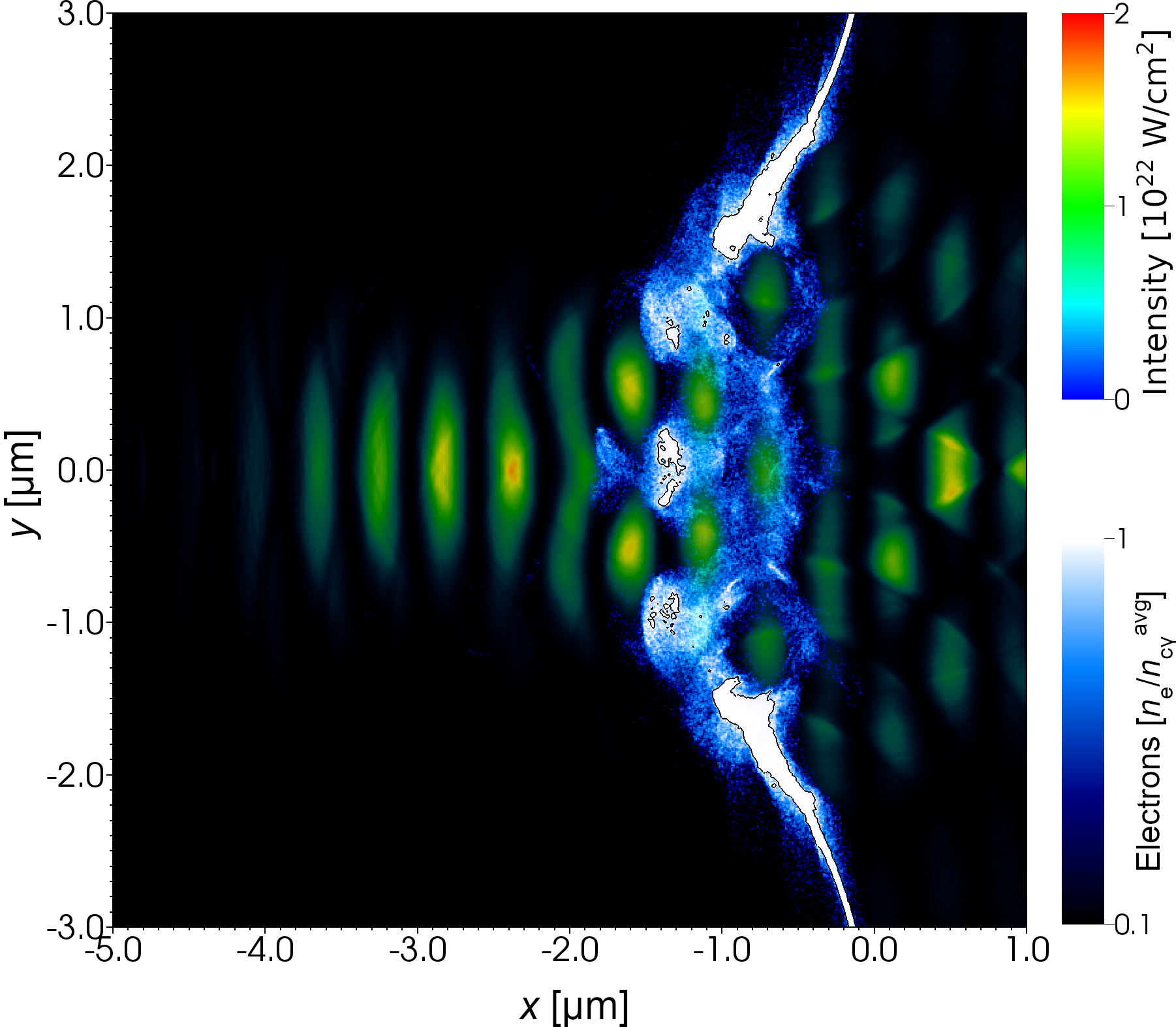}
\caption{The intensity of the s-polarized laser pulse and the density of target electrons for $ I_{0}=10^{22}~\mathrm{W/cm^{2}} $ and $ \lambda=805~\mathrm{nm} $. Target thickness is 30 nm. Black contours in electron density represent the relativistic critical density $ n_{\mathrm{c}\gamma}^{\mathrm{avg}} $.}
\label{fig:s}
\end{figure}

\begin{figure}
\centering
\includegraphics[width=1.0\linewidth]{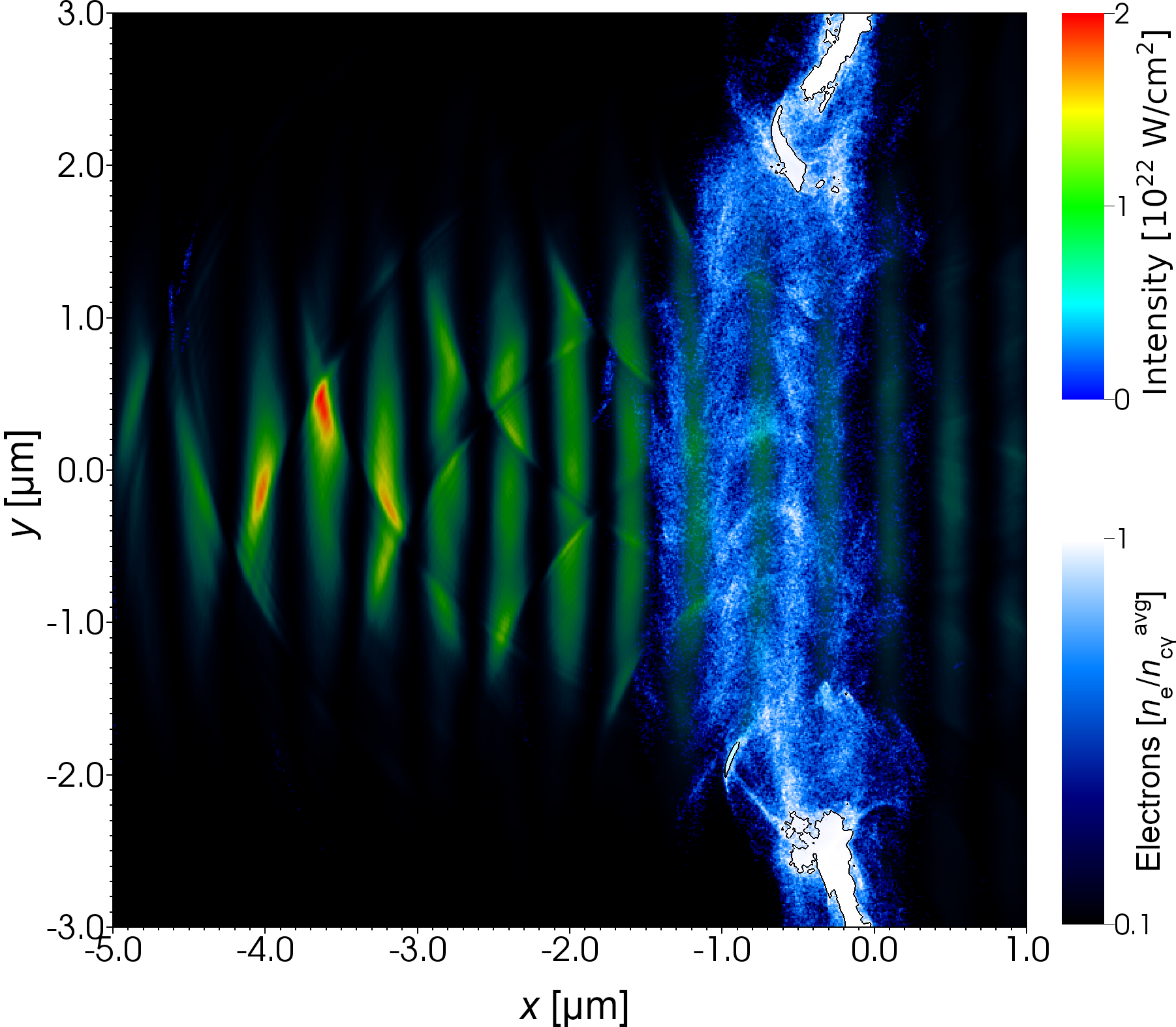}
\caption{The intensity of the p-polarized laser pulse and the density of target electrons for $ I_{0}=10^{22}~\mathrm{W/cm^{2}} $ and $ \lambda=805~\mathrm{nm} $. Target thickness is 30 nm. Black contours in electron density represent the relativistic critical density $ n_{\mathrm{c}\gamma}^{\mathrm{avg}} $.}
\label{fig:p}
\end{figure}

However, in both cases, the field amplification is expected to be limited by the target thickness that has to be smaller than the relativistically corrected skin depth $ l_{\mathrm{skin}}$.
Since Eq.~\eqref{a0reqt} is maximized at $ l= l_{\mathrm{skin}} $ for given target density and laser wavelength, such a choice presents through Eqs.~\eqref{a0resS} and ~\eqref{a0resP} the upper limit for the resulting field strength.

The longitudinal position of the maximum field strength at the rear side of the target can be estimated from the knowledge of the aperture size and propagation direction of the transmitted or scattered light, see Fig.~\ref{fig:schematic}.
The minimal aperture radius of the foil can be estimated as
\begin{equation}\label{w0Pinit}
w^\mathrm{min}=w_{0}\sqrt{\left|\ln\dfrac{a_{0}^{\mathrm{max}}}{a_{0}}\right| },
\end{equation}
as illustrated in Fig.~\ref{fig:schematic}(c).

In the case of s-polarization, the target surface is perturbed by the action of created electromagnetic nodes.
The propagation direction $ \theta $ (with respect to the laser axis) of the transmitted beam at the rear side of the target is given by the deformation of the target surface caused by radiation pressure and, therefore, it strongly depends on laser and target parameters \cite{GonzalezIzquierdo2018Radiation,McIlvenny2020}.
However,  $ \tan\theta $ can be estimated as the ratio of the laser pulse strengths at the axis and at the radial distance which are sufficiently strong to penetrate the target.
Below the strength limit required for target penetration, the foil deformation is given by the distribution of intensity iso-surfaces of the incoming laser beam.
In such a case, the angle can be estimated as $ \tan\theta= \tau c/\left( \sqrt{2\ln2}w_{0}\right)$.
When the plasma aperture is created, the ratio of longitudinal and radial distances given by iso-surfaces has to be weighted by the maximum $ a_{0}^{\mathrm{max}} $ and the characteristic value $ a_{0}^{\mathrm{t}} $ of field strengths required for penetration which results in
\begin{equation}\label{tanTheta}
\tan\theta=\dfrac{\tau c}{\sqrt{2\ln2}w_{0}}\dfrac{w^\mathrm{min}}{w^{\mathrm{avg}}},
\end{equation}
where
\begin{equation}\label{w0Pavg}
w^\mathrm{avg}=w_{0}\sqrt{\left|\ln\dfrac{a_{0}^{\mathrm{t}}}{a_{0}}\right| }.
\end{equation}
is considered, see Fig.~\ref{fig:schematic}(d).
Assuming, that the laser pulse starts to penetrate once $ a_{0}^\mathrm{max} $ is achieved, one can expect the interference maxima to occur at the distance
\begin{equation}\label{fs}
f_{\mathrm{s}}=w^\mathrm{min}\tan\theta
\end{equation}
from the initial target position.

In the case of p-polarization, the maximum field strength is created by the constructive interference of scattered harmonics with the field of the propagating laser beam.
Thus its location depends on the propagation direction of scattered photons which is a function of electron beam momentum.
However, the angle of the electron bunch propagation direction evolves with time.
At the beginning, the electrons are driven along the $ y $-axis.
At later times, the angle of propagation $ \psi $ can be estimated again from the geometrical properties of the Gaussian laser pulse as
\begin{equation}\label{tanPsi}
\tan\psi=\dfrac{\tau c}{\sqrt{2\ln2}w_{0}}.
\end{equation}
The laser photons are scattered into a number of harmonics, each having a different angle of propagation measured with respect to the electron beam direction.
Among them the second harmonic has a dominant contribution for the interference pattern and the corresponding angle can be estimated as \cite{Landau1980Classical} 
\begin{equation}\label{theta2nd}
\theta_{\mathrm{2nd}}=\arccos\left[  \dfrac{1+\beta\cos\left( \psi/2\right) }{2\beta}\right], 
\end{equation}
where $ \beta=\sqrt{1-1/\left( \gamma^{\mathrm{avg}}\right) ^{2}} $, $ \gamma^{\mathrm{avg}}=\sqrt{1+\left( a_{0}^{\mathrm{t}}\right) ^{2}/2}$ and $ \psi/2 $ is the average value of the angle at which the electron bunch propagates.
Thus, the maximum field strength is expected to be achieved at the distance of 
\begin{equation}\label{fp}
f_{\mathrm{p}}=w^{\mathrm{avg}}\tan\left( \theta_{\mathrm{2nd}}+\psi/2\right) .
\end{equation}

\section{2D and 3D simulation results}
We compare our theoretical predictions with 2D/3D PIC simulations performed in the code {\sf EPOCH} \cite{Ridgers2014,Arber2015}. 
The 2D simulation box resolved with 10 125 $ \times $ 4 500 cells is spanning from -30 $ \mathrm{\mu m} $ to 15 $\mathrm{\mu m} $ in the $ x $-direction and from -10 $\mathrm{\mu m} $ to  10 $\mathrm{\mu m} $ in the $ y $-direction.
The laser pulse of the peak intensity $ I_{0}=10^{22}~\mathrm{W/cm^{2}} $ ($ a_{0}=69 $) enters the simulation box at a boundary $ x=15~\mathrm{\mu m} $ and is focused to a spot of radius $ w_{0}=1.5~\mathrm{\mu m} $ located at $ x=0~\mathrm{\mu m} $.
It has a wavelength $ \lambda=805~\mathrm{nm} $ and a Gaussian temporal envelope of FWHM duration $ \tau=30~\mathrm{fs}$ in laser intensity.
We performed simulations for both s- and p-polarization of the laser pulse.
The $ \mathrm{Al}^{13+} $  solid foil with the corresponding free electron density 450$ n_{\mathrm{c}} $ and the thickness $ 10-60~\mathrm{nm}  $ is located at the focal spot.

The distributions of the transmitted laser intensity and the density of target electrons are shown in Figs.~\ref{fig:s} and \ref{fig:p} for s- and p-polarization, respectively.
The black contours correspond to the relativistic critical plasma density $ n_{\mathrm{c}\gamma}^{\mathrm{avg}}=\gamma^{\mathrm{avg}}n_{\mathrm{c}}  $.
In the first figure it is shown, that during the penetration of the target having a thickness of $ 30~\mathrm{nm} $, the laser field has a two-lobe structure (at $ x=-1.6~\mathrm{\mu m} $) creating later a symmetric field distribution at $ x=-2.4~\mathrm{\mu m} $.
This is a significant difference from the latter case, in which an asymmetric field diffraction pattern is created due to the interference of the penetrating laser field with the field scattered by the dense electron bunches which are driven alternately from two opposite sides of the aperture (see contours of the critical density near $ x= -0.6~\mathrm{\mu m}$, $ y=2~\mathrm{\mu m} $ and $ x= -1~\mathrm{\mu m}$, $ y=-2~\mathrm{\mu m} $ in Fig.~\ref{fig:p}).
%
\begin{figure}
\centering
\includegraphics[width=1.0\linewidth]{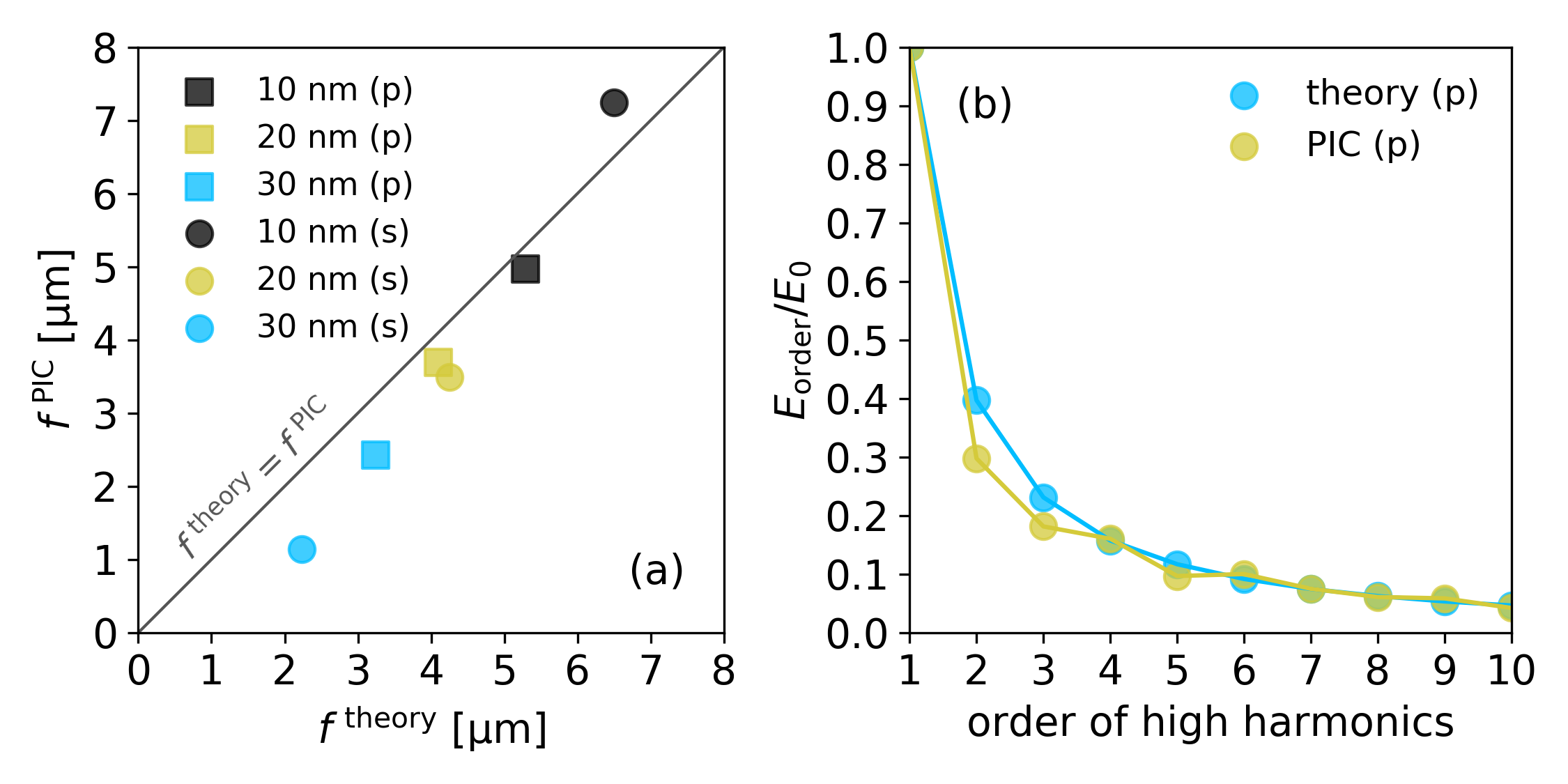}
\caption{(a) Longitudinal position of the maximum field strength at the rear side of the target given by Eqs.~\eqref{fs} and \eqref{fp} and obtained from PIC simulation. (b) The ratio of the first ten harmonics of the scattered light to the incident field of strength $ E_{0} $ obtained from PIC simulations and compared to the theory described in Refs.~\cite{Baeva2006,Edwards2020} for p-polarized laser beam.}
\label{fig:orderandfocus}
\end{figure}
\begin{figure}
\centering
\includegraphics[width=1.0\linewidth]{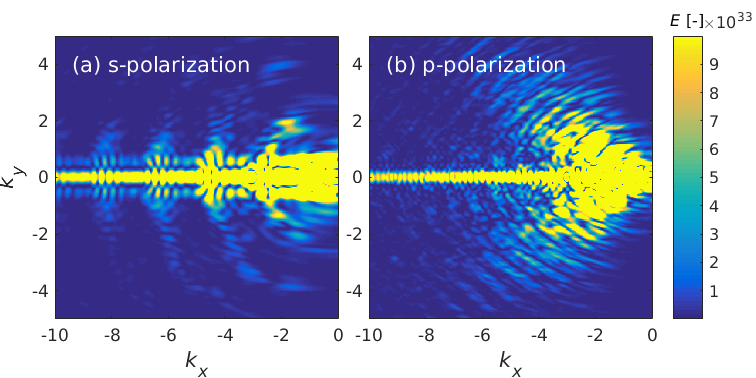}
\caption{The electric field intensity distribution in $ k_{x}-k_{y} $ phase-space for (a) s- and (b) p-polarization. The laser of intensity $ I_{0}=10^{22}~\mathrm{W/cm^{2}} $ and wavelength $ \lambda=805~\mathrm{nm} $ irradiates the 30 nm thick target. The color scale is saturated to enhance the visibility of the higher frequency components.}
\label{fig:test}
\end{figure}

The longitudinal position of the maximum field strength generated by the interference pattern at the rear side of the target is shown in Fig.~\ref{fig:orderandfocus}(a).
The expected values calculated from Eqs.~\eqref{fs} and \eqref{fp} ($ x $-axis) are compared to the results obtained from PIC simulations ($ y $-axis) for different target thicknesses.
The solid line is added to guide the eye for the perfect agreement between the theory and simulation results.
As expected according to the theory, the focal distance becomes shorter as the target thickness grows.

In Fig.~\ref{fig:orderandfocus}(b) we compare the ratio of the first ten harmonics of the scattered light to the incident field of amplitude $ E_{0} $ in case of p-polarized laser beam.
The respective harmonics are obtained by filtering out all other frequency components in the Fourier transform of the laser electric field.
The results obtained from the simulation with $ 30~\mathrm{nm} $ target are in good agreement with the theory described in Ref.~\cite{Baeva2006,Edwards2020} which was used for calculating the resulting field amplitude $ a_{0}^{\mathrm{p}} $ in Eq.~\eqref{a0resP}.
It can be seen that the field strength for higher harmonics is small and it can be neglected. 
This justifies our choice of $ N=10 $ in Eq.~\eqref{a0resP}.
The electric field distributions in $ k_{x}-k_{y} $ phase-space ($ \mathbf{k} $ is the wave-vector) for (a) s- and (b) p-polarization are shown in Fig.~\ref{fig:test} for the same laser and target parameters.
This spectrum is calculated only for the field which penetrates through the target.
There is a clear evidence of high harmonics propagating at an angle in case of p-polarization in contrast to s-polarization.
The $ k_{x} $ spectrum is broad as the maximum intensity due to interference is very well localized corresponding to a short spatial profile of the pulse in the longitudinal direction.
The $ k_{x} $ and $ k_{y} $ are normalized to the vacuum wavenumber of the laser $ k_{0}=2\pi/\lambda $.

\begin{figure}
\centering
\includegraphics[width=1.0\linewidth]{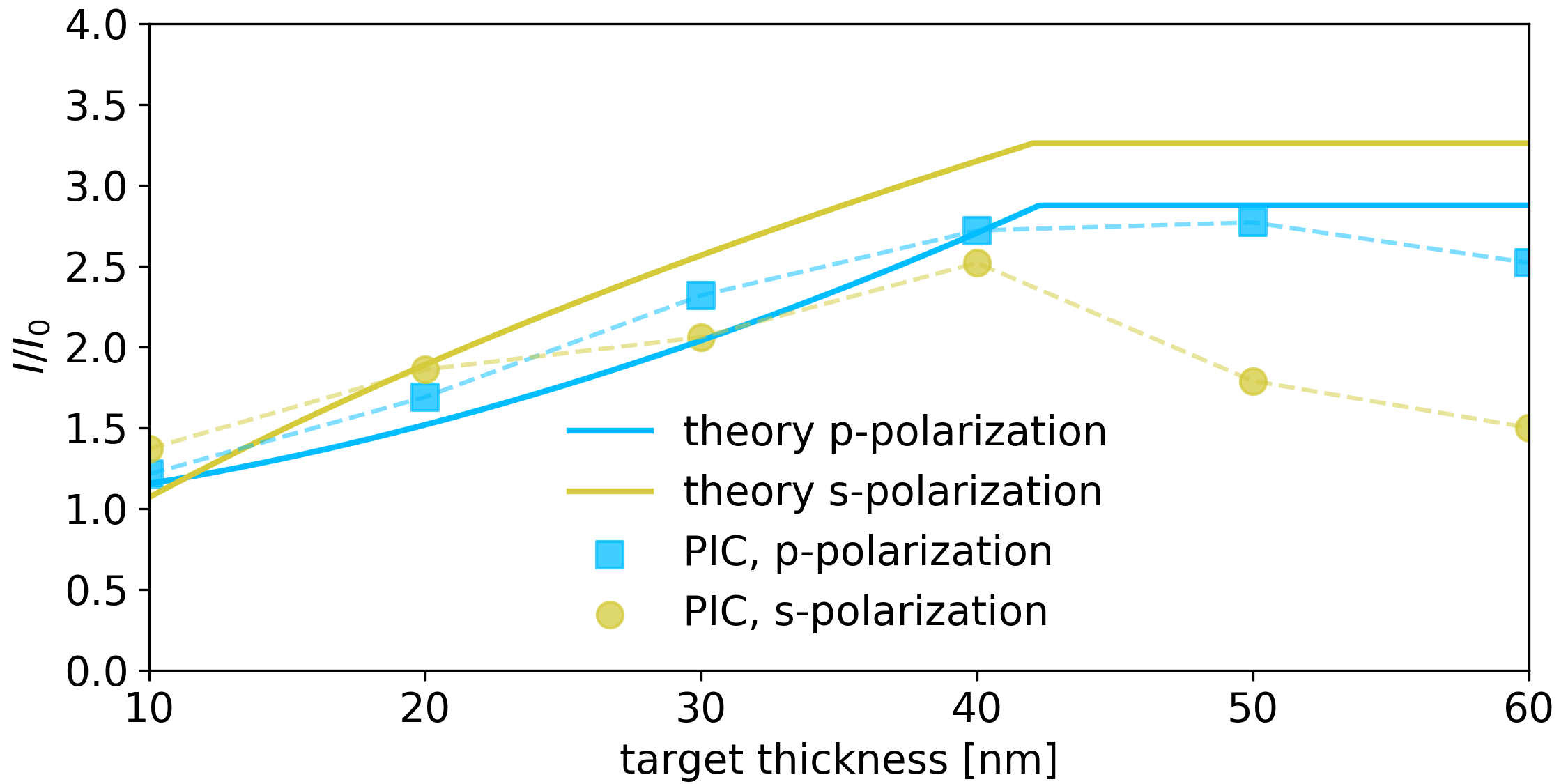}
\caption{Maximum intensity $ I $ achieved in the interaction of the s- and p-polarized Gaussian laser beam characterized by $ I_{0}=10^{22}~\mathrm{W/cm^{2}} $ with the target of different thickness. Lines represent the expected values given by Eqs.~\eqref{a0resS} and ~\eqref{a0resP}, markers show the results obtained from PIC simulations. The relativistically corrected skin depth is $ 42~\mathrm{nm} $.}
\label{fig:fig01}
\end{figure}

Figure \ref{fig:fig01} presents the comparison of resulting field amplitude as a function of target thickness for both types of a linearly polarized laser beam.
The simulation results show that the peak intensity of the diffracted field gradually grows with the target thickness as predicted by Eqs.~\eqref{a0resS} and Eqs.~\eqref{a0resP}.
The thicker the target is, the stronger field is required for penetration, and thus the resulting diffracted field achieves higher amplitude.
However, increasing the target thickness beyond the relativistically corrected skin depth $ l_{\mathrm{skin}}=42~\mathrm{nm} $ results in more pronounced laser energy absorption in the target and thus the intensity of the diffracted field becomes reduced.
The obtained results are therefore in good agreement with the theory as it  has a natural limit for target thickness given by the relativistically corrected skin depth.
This absorption is not taken into account in the theory and thus the theoretical curve stays constant.
In case of p-polarization, the first ten harmonics were considered in Eq.~\eqref{a0resP}.

In order to check the validity of our results and get more realistic information on the field amplification due to the interference of the laser field, we carried out an additional 3D simulation for different laser and target parameters ($ \lambda=1~\mathrm{\mu m} $, $ \sin^{2} $ temporal envelope, foil thickness 20~nm, $ n_{\mathrm{p}}=835n_{\mathrm{c}} $).
The simulation box of dimensions $ 44~\mathrm{\mu}m\times17~\mathrm{\mu}m\times17~\mathrm{\mu}m $ was resolved with $ 8~800\times680\times680 $ cells.
\begin{figure}
\centering
\includegraphics[width=1.0\linewidth]{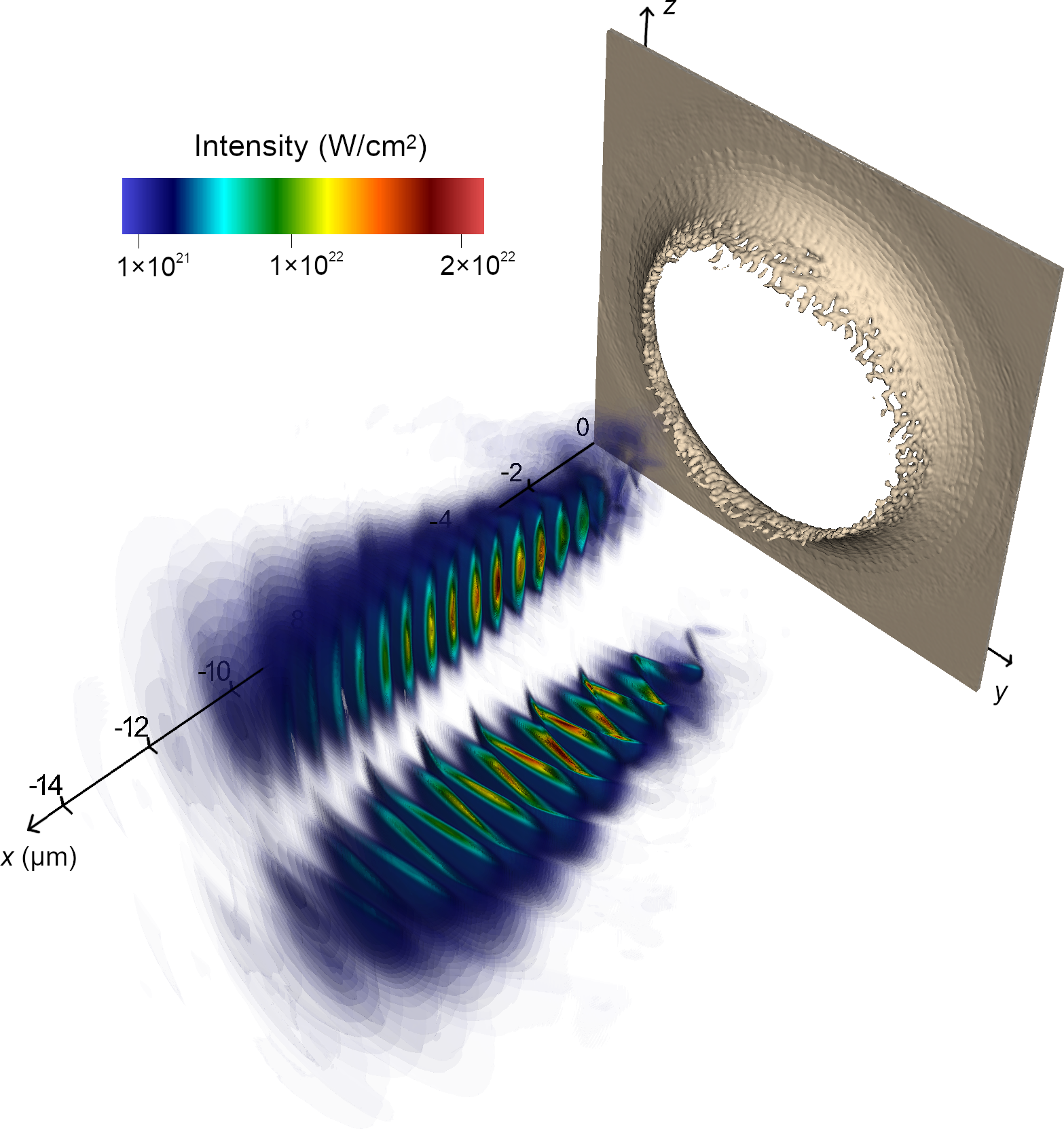}
\caption{The distribution of the laser intensity in horizontal and vertical slices of the laser pulse after the interaction with a foil as obtained from 3D simulation.}
\label{fig:puls}
\end{figure}
As shown in Fig.~\ref{fig:puls}, the characteristic structures of the diffracted laser field in a plane perpendicular or parallel to the polarization axis (see Figs.~\ref{fig:s} and ~\ref{fig:p}) are present also in 3D case.
The horizontal and vertical slices of the laser beam correspond to p- and s-polarization from 2D, respectively.
The horizontal plane in 3D contains the electric field vector of the linearly polarized laser beam.
The laser field amplification due to the diffraction on an evolving plasma aperture in 3D case can be roughly estimated as the product of contributions from two planes which are perpendicular ('s') and parallel ('p') to the polarization direction and which were studied above in 2D case.
This approach predicts the maximum amplification of the laser intensity by a factor of 7.0 at a distance $ 3.7~\mathrm{\mu}m $ behind the shutter, which is comparable to the maximum value 7.1 obtained from 3D simulation at a distance $ 4.3~\mathrm{\mu}m $, see Fig.~\ref{fig:test2}, having in mind that in the simulation it is not guaranteed that the field interference maxima in both planes appear simultaneously at the same position.

\begin{figure}
\centering
\includegraphics[width=1.0\linewidth]{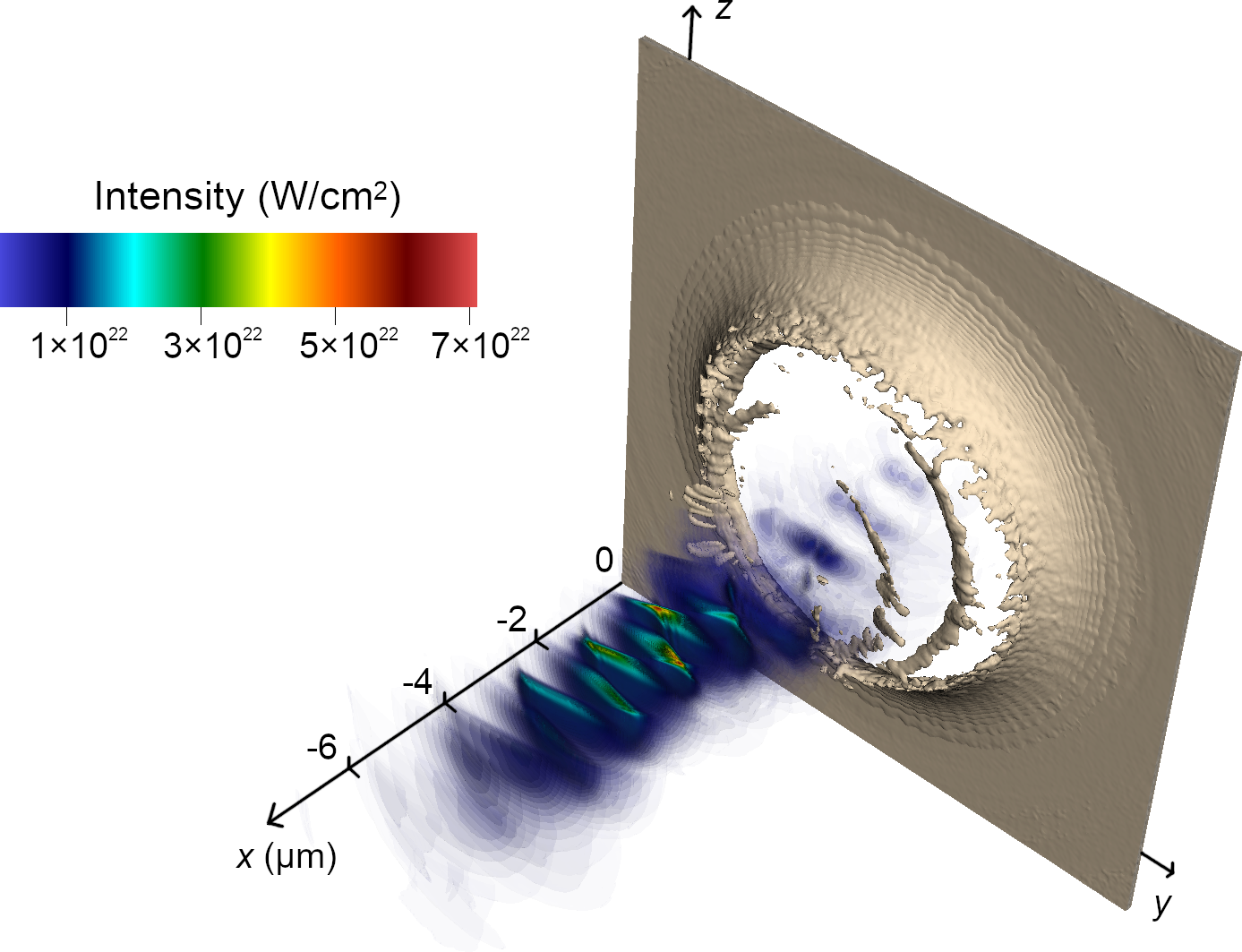}
\caption{The distribution of the laser intensity in the horizontal slice of the laser pulse as obtained from 3D simulation at the moment of reaching the maximum value.}
\label{fig:test2}
\end{figure}

So far we have assumed an idealized laser pulse.
However, the laser systems are subject to a limited contrast \cite{Weber2017}.
Due to the interaction of the laser pulse pedestal with the target, the pre-plasma is formed before the main laser pulse arrives and thus it can significantly affect the interaction dynamics.
Considering the above mentioned laser intensity, the pedestal of the main laser pulse can completely destroy the nm-scale foil.
%
%
%
%
%
%
%
%
%
%
The foil expands to several microns along the propagation axis and therefore its density is below the relativistically corrected critical value for the main laser pulse.
As a result, the main laser pulse propagates in the expanded plasma without any significant distortion or focusing.
To avoid such a scenario in a real experiment, one might consider the interaction with two foils.
The pedestal of the main laser pulse interacts with the first foil only.
The main laser pulse then propagates trough the created under-dense plasma and interacts with the second foil.
Such a setup allows creating a relativistic plasma aperture in the second foil which is essential for laser pulse diffraction resulting in the increase of the laser intensity \cite{Nikl2021} which can be used, e.g., for enhancement of ion acceleration from an additional target \cite{Matys2021}.
%


\section{Conclusion}
In conclusion, we have studied the process of laser pulse diffraction on the relativistic plasma aperture created in the interaction with an over-dense ultra-thin foil.
Such a setup allows amplifying the laser intensity due to the interference of the transmitted/diffracted beam.
We provide the theoretical estimates for obtaining the maximum field strength and its spatial location as a function of the laser and target parameters.
Using 3D numerical simulations we have verified that the laser intensity can be increased at least seven-times when proper laser and target parameters are used.
This setup, based solely on relativistic plasma optics, thus presents a viable approach for obtaining localized laser intensity enhancement which might be interesting for applications like ion acceleration, generation of gamma rays and electron positron pair creation.
%


\section*{Acknowledgements}
Portions of this research were carried out at ELI Beamlines, a European user 
facility operated by the Institute of Physics of the Academy of Sciences of the 
Czech Republic.
This work is supported by the project High Field Initiative (HIFI) CZ.02.1.01/0.0/0.0/15\_003/0000449 from European
Regional Development Fund (ERDF). The support of Czech Science Foundation project No. 18-09560S is acknowledged. The support of Grant Agency of the Czech Technical University in Prague is appreciated, grant no. SGS19/192/OHK4/3T/14. This work was supported by The Ministry of Education, Youth and Sports from the Large Infrastructures for Research, Experimental Development and Innovations project "IT4Innovations National Supercomputing Center – LM2018140". Computational resources were supplied by the project "e-Infrastruktura CZ" (e-INFRA LM2018140) provided within the program Projects of Large Research, Development and Innovations Infrastructures.
The simulations were performed at the cluster ECLIPSE at ELI Beamlines.

\bibliography{shutter}{}
\bibliographystyle{apsrev4-2}

%


\end{document}